\documentclass[twocolumn,showpacs,preprintnumbers,amsmath,amssymb]{revtex4}

\usepackage{graphicx}% Include figure files
\usepackage{dcolumn} % Align table columns on decimal point
\usepackage{bm}      % bold math

\usepackage{textcomp}

\begin{document}

\preprint{APS/123-QED}

\title{Limitations for the determination of piezoelectric constants \\
       with piezoresponse force microscopy}

\author{Tobias Jungk}
\email{jungk@uni-bonn.de}
\author{\'{A}kos Hoffmann}
\author{Elisabeth Soergel}

\affiliation{Institute of Physics, University of Bonn,
Wegelerstra\ss e 8, 53115 Bonn, Germany\\}

\date{\today}

\begin{abstract}
%PRB: 5% vom Text aber < 500 Worte%
%APL: < 100 Worte%
%
At first sight piezoresponse force microscopy (PFM) seems an ideal
technique for the determination of piezoelectric coefficients (PCs),
thus making use of its ultra-high vertical resolution ($
<$~0.1\,pm/V).
Christman et al.~\cite{Chr98} first used PFM for this purpose.
Their measurements, however, yielded only reasonable results of
unsatisfactory accuracy, amongst others caused by an incorrect
calibration of the setup.
In this contribution a reliable calibration procedure is given
followed by a careful analysis of the encounted difficulties
determining PCs with PFM.
We point out different approaches for their solution and expose why,
without an extensive effort, those difficulties can not be
circumvented.
\end{abstract}


\maketitle

The determination of the magnitude of the piezoelectric coefficients
(PCs) is still a rewarding experimental challenge. Although several
methods  have been utilized (measuring the resonance frequencies of
specifically cut samples \cite{Ogi02}, determining the velocity of
sound \cite{Kov90}, utilizing a Berlincourt meter \cite{Zha97} or
using optical heterodyne interferometry \cite{Roy92}), they all
suffer from being cumbersome and the published values vary strongly.
In the past years piezoresponse force microscopy (PFM) has become a
standard tool for investigating ferroelectric and thus piezoelectric
samples \cite{New} making use of the converse piezoelectric effect.
A detailed description of PFM can be found elsewhere
\cite{Alexe,Jun06a}. However, despite its ultra-high vertical
resolution in the sub-picometer regime, it is not applied for the
precise determination of PCs. This is mainly due to the fact, that
even well established PCs could not be reliably confirmed with PFM.
Interestingly, the failure of PFM measurements with high
quantitative accuracy is mostly due to the incorrect calibration the
instrument. However, even with appropriate calibration, PFM is not
capable of determining piezoelectric coefficients.

In this contribution, we will show why the calibration technique
generally referred to~\cite{Chr98} leads to wrong PFM-calibration
constants. We will in return present a reliable calibration
procedure.
We will further more focus on the difficulties to determine PCs with
high accuracy with PFM. It will unfortunately turn-out that a
precise determination of PCs with PFM is generally not possible.

Piezoresponse force microscopy is based on the deformation of the
sample due to the converse piezoelectric effect. The PFM is a
scanning force microscope (SFM) operated in contact mode with an
alternating voltage $U_{\rm tip}$ applied to the tip. In
piezoelectric samples this voltage causes thickness changes $\Delta
t$ and therefore vibrations of the surface which lead to
oscillations of the cantilever that can be read out with a lock-in
amplifier.
In order to obviate misunderstandings we briefly define the symbols
used later on for the outputs of the lock-in amplifier (LI). A
LI-signal can generally be described in a circular coordinate system
as a vector $\bf P$ with an appropriate length $P$ and an angle
$\theta$ with respect to the reference signal. This is the way the
signals are read-out from a single-phase LI.
In case of a dual-phase LI, the two output signals can optionally
also be displayed in a cartesian coordinate system thus resulting in
the output signals  $P^X$ and $P^Y$.
Note that adjusting the phase $\phi$ of the reference signal
corresponds to a rotation of the coordinate system.
In the following we will adapt the usual notation naming the output
signals $P$ and $\theta$ ($\phi = 0$) of the LI the PFM-signals,
subscripts specify the particular sample.

In general the calibration of the SFM for PFM-measurements is
performed according to the hence often cited work by Christman et
al.~\cite{Chr98}. In brief, a piezoelectric sample with known PC is
brought into the PFM-setup. Due to its very precisely determined PC
an $\alpha$-quartz-plate ($d_{11}=2.3\pm 0.05$\,pm/V~\cite{Lan}) is
generally used for this purpose. One presumes thus to obtain the
relation between the PFM-signal $P_{\alpha}$ and the vibration
amplitude $\Delta t = d_{11}\,U_{\rm tip}$ of the surface, leading
to a PFM-calibration constant $k_{\rm PFM} = \Delta t /P_{\alpha}=
d_{11} \, U_{\rm tip} / P_{\alpha}$.
Thus by measuring $P$ of any other sample one presumes to determine
its particular PC as
\begin{equation}\label{eg:Jungk1}
 d = k_{\rm PFM} \, P / U_{\rm tip} = d_{11} P / P_{\alpha} \quad .
\end{equation}

However, due to the system-inherent background \cite{Jun06a} this
calibration procedure and thus the precise determination of PCs
fails. This background, present in current SFM setups, shows-up as a
frequency dependent contribution to the PFM-signal, probably caused
by a wealth of mechanical resonances of the SFM-head. The amplitude
of the background (1--10\,pm/V) scales linearly with $U_{\rm tip}$,
its phase varies randomly. Possible consequences of the
background-adulterated PFM-signals have been presented in detail
elsewhere \cite{Jun07b}. Hitherto attempts to suppress the
background by modification of the SFM-head failed.

\begin{figure}
\includegraphics{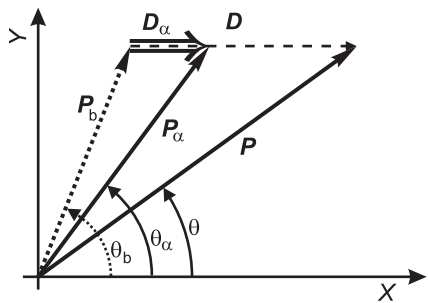}
\caption{\label{fig:Jungk1}
Output signals of the lock-in amplifier in the $X$-$Y$-plane. The
system-inherent background is ${\bf P}_b$.
A piezoelectric standard sample with a piezoelectric coefficient
$d_{\alpha}$ contributes ${\bf D}_{\alpha}$ $(\Rightarrow)$ to the
resulting PFM-signal $\bf P_{\alpha}$. Same applies for the sample
with the PC to be measured: it's contribution to the measured signal
$\bf P$ is $\bf D$ $(\dashrightarrow)$.
} \end{figure}

Figure~\ref{fig:Jungk1} depicts the situation showing PFM-signals in
a vectorial diagram. Note that $\bf D_{\alpha}$ denotes the
contribution of the calibration sample to the PFM-signal which
should not be confused with its piezoelectric coefficient
$d_{\alpha}$. For the sample to be measured the same notation ($\bf
D$ and $d$) applies.
Starting from scratch: the background ${\bf P}_b$ results in
PFM-signals of $P_b$ and $\theta _b$.
The contribution of the calibration sample adds on the background as
${\bf P}_{\alpha} = {\bf P}_b + \bf D_{\alpha}$ thus resulting in
the PFM-signals $P_{\alpha}$ and $\theta_{\alpha}$.
The sample to be measured finally leads to the PFM-signals $P$ and
$\theta$. From simple geometrical considerations is can be seen that
\begin{equation}\label{eg:Jungk2}
P_{\alpha}
=\sqrt{P_b^2+D_{\alpha}^2+2P_bD_{\alpha}\cos\theta_b}\quad .
\end{equation}
The same expression applies for $P$ substituting $D_{\alpha}$ by
$D$.
It is now self-evident that because ${\bf P}_b\neq 0$ and thus $\bf
P_{\alpha}\neq D_{\alpha}$ the procedure described previously for
calibrating the PFM and consequently the determination of $d$ fails.

An example shows the significance of the situation described above.
In Fig.~\ref{fig:Jungk1} the different contributions to the
PFM-signals are shown in realistic proportions when using
$\alpha$-quartz ($d_{11}=2.3$\,pm/V) for calibration. Let us assume
the background to have an amplitude of $P_b = 8$\,pm/V at an angle
of $\theta_b=60^{\circ}$ and the sample to have a PC of $d=7$\,pm/V.
Using Eqs~\ref{eg:Jungk1} and \ref{eg:Jungk2} would lead to $d =
d_{11}\, P / P_{\alpha} \simeq 3.2$\,pm/V which is wrong by a factor
of more than two.
%

%The Calibration

To overcome the above mentioned difficulties it is essential to
conduct a reliable calibration of the PFM. Therefore three steps
have to be accomplished:
\begin{itemize}
    \item Calibration of the z-scanner of the SFM. This can be
    accomplished with a height standard.
    \item Determination of thickness change $\Delta t$ of the
    calibration sample at a specific voltage $U_{\rm tip}$ and
    frequency $f$ applied to the tip. This measurement is performed
    with the ''height-modus'' of the SFM.
    Therefore a piezoceramic sample with a large
    PC ($\sim 500$\,pm/V) is most appropriate. The frequency $f$ must be
    low with respect to the feedback loop of the SFM thus
    the z-scanner can fully follow the movement ($f \sim 0.1$ -- 1\,kHz).

    The PC of the calibration sample has to be large in order to yield
    measurable thickness changes $\Delta t$ at moderate voltages.
    Applying e.g. $10\,$V$_{\rm pp}$ to the tip for a sample with a PC of 500\,pm/V results in a
    thickness change of $\Delta t_{\rm PZT}=5\,$nm, easily detectable with SFM.
    Note that $\alpha$-quartz is not suited as it would only give
    $\Delta t=$0.023\,nm. This, however, is not measurable
    with the ''height-modus'' of the SFM.
    \item With the piezoceramic sample used in the step before,
    using the same voltage and the same frequency applied to the tip
    the PFM can now be calibrated just by disabeling the feedback-loop
    and measuring the output $P_{\rm PZT}$ of the lock-in amplifier.
    One thus gets the wanted calibration constant
    $k_{\rm PFM} = \Delta t_{\rm PZT} / P_{\rm PZT}$.
\end{itemize}
Another possibility to calibrate the SFM by determining the detector
sensitivity via a force-distance measurement and then calculating
the expected PFM-signal \cite{Har}. This calibration method has a
disadvantage since it is a low-frequency measurement whereas for PFM
usually frequencies in the kHz regime are used. It is thus not
performed under the same conditions than the PFM measurements.

%The background

\begin{figure}
\includegraphics{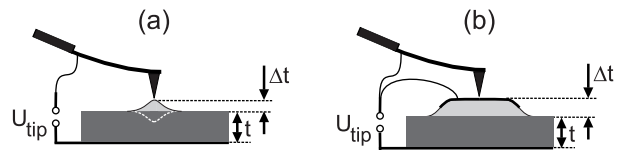}
\caption{\label{fig:Jungk2}
(a) The tip acting as electrode. Only a very small volume of the
sample (some\,\textmu m$^3$) contributes to the deformation.
(b) A large area top electrode. Beneath the central part of the
electrode, the whole volume of the sample expands homogeneously.
Near the edges clamping influences the deformation.
} \end{figure}

Now, after calibration of the PFM,  the main problem for determining
reliable PCs with the PFM remains the background. Although the
origin of the background is not fully understood, it can be
attributed to resonances of the whole SFM setup, i.e. head of the
microscope, and sample \& sample stage, depending on the specific
realization of the top electrode. Therefore two situations have to
be discussed individually: (a) the tip and (b) a large area
metallization acting as top electrode. In both cases the backside of
the sample is covered with a large area electrode
(Fig.~\ref{fig:Jungk2}).

{\sl Tip as electrode}. This is the standard configuration for
PFM-measurements yielding a high lateral resolution. In this
situation, the electric field inside the crystal can reach values of
up to $E \approx 10^9$\,V/m just underneath the tip, depending on
the tip radius and the dielectric constant of the sample. However,
due to the strong inhomogeneity caused by the sharp tip, $E$ decays
within $\leq 1\,\mu$m inside the sample \cite{Ott04}. The
piezoelectrically excited region is thus only a few \textmu$\rm
m^3$(Fig.~\ref{fig:Jungk2}(a)). This has two important consequences
on PFM-imaging: (i)~the whole sample is at rest, the background is
only due to the SFM-head (incl. cantilever) and can thus be
determined e.g.~with a glass plate~\cite{Jun06a}; (ii)~Due to
clamping of the surrounding material the crystals deformation is
drastically reduced. The values measured in this way were found to
be too small by a factor of up to three~\cite{Jun07a}.
As a result measurements with the tip acting as electrode are not
suited for the determination of PCs with high accuracy.

{\sl Large area electrode}. To avoid the problem of clamping and to
minimize any effect of electrostatic interaction between the
cantilever and the sample, a large area electrode of some mm$^2$ is
evaporated on top of the sample. Electrical contact is performed
directly to the electrode through an external wire, short-circuited
with the tip (Fig.~\ref{fig:Jungk2}(b)). Besides being unclamped in
the center of the electrode, this configuration offers another
advantage in comparison to the situation with the tip acting as
electrode: The electrical field applied to the crystal is well
defined as it is homogeneous across the whole sample thickness. In
this configuration, however, the background is no longer independent
of the sample which turns out to be a probably irresolvable
drawback. Since the whole sample is piezoelectrically excited,
sample and sample-holder also contribute to the background. A series
of experiments with different fixations of the sample (sticking it
with epoxy on a large lead block, embedding it in rubber, or
suspending it freely) failed. Every mounting showed its own
frequency dependence of the PFM-signal why a determination of the
background is not possible.
Thus, measurements with the large area electrodes are not suited for
the determination of PCs with high accuracy.

%Possible Solutions

Of course the immediate question arises whether there is any
possibility to circumvent the drawbacks presented above, thus
enabling a precise determination of piezoelectric coefficients. A
reliable measurement of PCs with the tip acting as electrode can be
excluded as it fails due to clamping, i.e., a fundamental physical
reason. For the measurements using large top electrodes, however,
the situation is different since so far technical deficiencies cause
the failure. In the following we will discuss three approaches that
might overcome the difficulties mentioned above.

(1) In principle, one can think about a SFM-setup suppressing any
kind of background. Although difficult to realize, it is not
impossible. As a reminder the amplitude of the background is of the
order of 10\,pm/V, which, using standard PFM settings (10\,V applied
to the tip) becomes comparable to the radius of an atom. The same
scale applies for the PCs to be measured. Obviously the specific SFM
we used (SMENA from NT-MDT) is not of low quality but lock-in
detection is amazingly sensitive. Note that in order to become
interesting for the determination of PCs, the background needs to be
reduced at least by two (!) orders of magnitude. Taking this into
account, building a ''background-free'' SFM appears as a remarkably
serious challenge.

(2) Another approach to circumvent the background-problem is based
on multi-domain samples. Measuring the PFM-signals on both domain
faces (with the same background) would automatically yield the
correct PC of the sample~\cite{Jun06a}. Apparently straightforward,
this approach results in new troubles: the clamping between adjacent
domains. As expected from theoretical considerations, and verified
experimentally~\cite{Jun07c}, the surface deformation is affected on
a length scale similar to the thickness of the sample. Thus, for a
500\,\textmu m thick sample, reliable values for the piezomechanical
deformation can only be obtained at a distance of $> 500$\,\textmu m
from any domain boundary. As a first consequence, the use of a large
bi-domain sample is required. Furthermore, the sample has to be
transferred by more than 1\,mm for measuring reasonable PFM-signals
on both domain faces. Unfortunately, when performing such crucial
changes in the mechanical setup, the background can not be presumed
to remain unchanged. This, however, is an absolute condition for the
quantitative analysis of a multi-domain measurement as proposed
here. Even worse, there is no way to find-out, whether the
translation of the sample did affect the background or not.

One could think, of course, to reduce this difficulty using thinner
crystals, e.g., 50\,\textmu m thickness, thus scaling the problems
described above by one order of magnitude.  But also a 100\,\textmu
m translation is still too much. Since for this measurement, the
samples must be free-standing, to make them even thinner is not
trivial. Thus although seemingly easy, this approach also suffers a
series of drawbacks not yet resolved.

(3) Finally, and so far the only realizable solution to the problems
described above consists of using very particular settings for the
PFM, avoiding the contribution of the background by applying
frequencies of $\ll 1$\,Hz to the tip, thus excluding the resonances
of the setup. This requires of course to disable the feedback of the
SFM. Due to the very long integration time of several minutes,
necessary for obtaining reliable data, those measurements can only
be carried out when the environment of the setup is at quiet as
possible. Although this measurement scheme might provide the best
data for PCs when using PFM, it is not a detection method suitable
to provide reliable values of high accuracy due to inherent drift
and noise in current SFM setups.

In this contribution, we extensively discussed the capability of
piezoresponse force microscopy (PFM) to provide accurate values for
the piezoelectric coefficients (PCs) of single crystals. We first
presented a calibration procedure for PFM measurements. In the
following we carried out a careful analysis of the different
measurement techniques and the belonging drawbacks. It turned out
that without an extensive effort PFM is not suited to yield reliable
PCs.

{\bf Acknowledgments}

We thank David Scrymgeour for helpful discussions. Financial support
of the DFG research unit 557 and of the Deutsche Telekom AG is
gratefully acknowledged.

%\clearpage


\begin{thebibliography}{15}
%1
\bibitem{Chr98}
J.~A.~Christman, J.~R.~R.~Woolcott, A.~I.~Kingon, and
R.~J.~Nemanich, Appl. Phys. Lett. \textbf{73}, 3851 (1998)

%2
\bibitem{Ogi02}
H. Ogi, Y. Kawasaki, M. Hirao, and H. Ledbetter,
J. Appl. Phys. \textbf{92}, 2451(2002).

%3
\bibitem{Kov90}
G. Kovacs, M. Anhorn, H. E. Engan, G. Visintini, and C. C. W. Ruppel,
IEEE Ultrasononics Symposium 435 (1990).

%4
\bibitem{Zha97}
Z.~Zhao, H.~L.~Chan, and C.~L.~Choy,
Ferroelectrics \textbf{195}, 35 (1997).

%4
\bibitem{Roy92}
D. Royer and V. Kmetik,
Electron. Lett. \textbf{28}, 1828 (1992).

%5
\bibitem{New}
R.~E.~Newnham {\it Properties of Materials} (Oxford University
Press, New York, 2005)

%6
\bibitem{Alexe}
M.~Alexe and A.~Gruverman, eds., {\it Nanoscale Characterisation of
Ferroelectric Materials} (Springer, Berlin; New York, 2004) 1st ed.

%7
\bibitem{Jun06a}%Quantitative analysis of ferroelectric domain imaging with piezoresponse force microscopy
T.~Jungk, A.~Hoffmann, and E.~Soergel,
Appl.\ Phys.\ Lett.\ \textbf{89}, 163507 (2006).

%8
\bibitem{Jun07b}%DU und der Lock-In
T.~Jungk, A.~Hoffmann, and E.~Soergel, submitted (2007).

%9
\bibitem{Lan}
Landolt-B{\"o}rnstein III/29 (Springer, Berlin; New York, 1979)

%10
\bibitem{Har}
C.~Harnagea and A.~Pignolet in {\it Nanoscale Characterisation of
Ferroelectric Materials, M.~Alexe and A.~Gruverman, eds.,} Chap.~2
(Springer, Berlin; New York, 2004) 1st ed.


%11
\bibitem{Ott04}
T.Otto, S. Grafstr{\"o}m, and L. Eng, Ferroelectrics \textbf{303},
149 (2004)

%12
\bibitem{Jun07a}%Influence of the inhomogeneous field at the tip on quantitative piezoresponse force microscopy
T.~Jungk, A.~Hoffmann, and E.~Soergel, Appl. Phys. A \textbf{86},
353 (2007).

%13
\bibitem{Jun07c}%Clamping
T.~Jungk, A.~Hoffmann, and E.~Soergel, submitted (2007).






\end{thebibliography}
\end{document}